\documentclass[prb,twocolumn,letterpaper]{revtex4}
\usepackage{graphicx,amsmath,color,ulem}

\begin{document} 

\title{Jaynes-Cummings treatment of superconducting resonators with dielectric loss due to two-level systems}
\author{M. Bhattacharya, K. D. Osborn, and Ari Mizel}
\affiliation{Laboratory for Physical Sciences, University of Maryland, College Park,
              MD 20740.}
\date{\today}
\pacs{PACS numbers}

\begin{abstract}
We perform a quantum mechanical analysis of superconducting resonators subject to dielectric loss arising from charged two-level systems. We present numerical and analytical descriptions of the dynamics of energy decay from the resonator within the Jaynes-Cummings model. Our analysis allows us to distinguish the strong and weak coupling regimes of the model and to describe within each regime cases where the two-level system is unsaturated or saturated. We find that the  quantum theory agrees with the classical model for weak coupling. However, for strong coupling the quantum theory predicts lower loss than the classical theory in the unsaturated regime. Also, in contrast to the classical theory, the photon number at which saturation occurs in the strong coupling quantum theory is independent of the coupling between the resonator and the two-level system.  
\end{abstract}

\maketitle

\section{Introduction}
Noise is a central issue in the implementation of quantum computation using
superconducting circuits \cite{MartinisDevoret2004,Shnirman2001}. Researchers have focused on two-level charge fluctuators \cite{Hunklinger1975,Arnold1976} as
candidate sources of energy- and coherence-loss in superconducting qubits \cite{Martinis20041, Martinis20042,Martinis2005, Nori2006,Makhlin2009,Martinis2008,Palmer2010}. Superconducting resonators, which are sensitive probes of bulk two-level system (TLS) loss, are also useful for dispersive qubit readout using circuit QED \cite{Girvin2004}, single artificial-atom lasing \cite{Nakamura2007} and single photon detection \cite{Day2003}. Experiments focusing on dielectric loss in superconducting resonators have previously employed a classical description of the resonator \cite{Martinis2009,Osborn2010}. However, many of these experiments involve low microwave-photon numbers and properly require a full quantum mechanical analysis for their interpretation \cite{Milburn2008,Shatokhin2008}. 

In this article we quantize the classical theory of loss due to two-level systems (TLSs) to obtain the familiar Jaynes-Cummings model (JCM) of quantum optics \cite{Cummings1963}.  Although the JCM has been studied thoroughly in both atomic and solid-state physics, to the best of our knowledge it has not been used to explore the loss regimes presented here \cite{Simmonds2007,Jacobs2009}, nor has it specifically been used to evaluate loss tangents. (In a similar way, there has been recent renewed interest in the JCM arising from nanofabricated solid state systems that are beginning to broach parameter regimes inaccessible to atomic physics experiments \cite{Solano2010,Girvin2010,Grifoni2010,Anirban2010}). 

In this paper we use the JCM to analyse a simplified set-up designed to capture the essence of resonator loss : a harmonic oscillator that can release quanta to a zero-temperature thermal bath only through a resonant TLS. We neglect the fact that a resonator may interact with many TLSs. Working with this minimal model allows us to establish the dynamics of decay and to demarcate various parameter regimes. This model can serve as a starting point for studies of quantum mechanical loss in more complex systems. It also allows us to generate transparent
analytic expressions for the resonator photon number, TLS excitation and coherence, as well as system correlations, which are of interest not only in the context of resonator loss but also for control and manipulation of the full system \cite{Nori2006,Jacobs2009}. Our analysis shows agreement between the quantum and classical theories for weak coupling between the resonator and the TLS. However in the case of strong coupling, we find the quantum theory predicts a significantly lower loss than the classical theory when the TLS is unsaturated. Also, in contrast to the classical case, the photon number at which the loss saturates does not depend on the resonator-TLS coupling. 

The remainder of the paper is arranged as follows. In Section~\ref{sec:CQ} we 
summarize the classical model of loss and proceed to quantize it, arriving at the
Jaynes-Cummings model; in Section~\ref{sec:NUM} we describe two numerical approaches towards solving the problem, using the master equation and the Maxwell-Bloch equations respectively; in Section~\ref{sec:AN} we present
approximate analytical solutions to the Maxwell-Bloch equations which allow us to
identify loss regimes exhibiting qualitatively different behavior. Section~\ref{sec:CONC} supplies a conclusion.

\section{Classical and quantum models of loss}
\label{sec:CQ}
In this section we briefly restate the TLS model that treats the superconducting resonator classically and proceed to quantize it to obtain the JCM. This exercise allows us to directly relate the parameters of the two models. 
\subsection{Classical model}
The classical model consists of a defect with charge $q$ that can 
tunnel between two spatially distinct sites separated by a distance $l$ inside the
dielectric that is permeated by the electric field of the superconducting resonator \cite{Phillips1987}. The onsite energies of the two states $|L \rangle$ and $|R\rangle$  are $\pm \Delta/2$ respectively and the height of the tunneling barrier is $\Delta_{0}$. The defect is driven by an electric field of magnitude $F(t)$ which is directed at an angle $\theta$ with respect to a line joining the two charge sites. The classical Hamiltonian of this system may be written as 
\begin{equation}
\label{eq:smclassK}
H_{C}=\left[\frac{E}{2}+\frac{p \cos \theta F(t) \Delta}{2E}\right]\sigma_{z}+\frac{p\cos \theta F(t)\Delta_{0}}{2E}\sigma_{x}+U,
\end{equation}
 where $E^{2}=\Delta^{2}+\Delta_{0}^{2}, p=ql$ and the Pauli matrices $\sigma_{z}$ and $\sigma_{x}$ have been defined in the energy basis
\begin{eqnarray}
|+\rangle &= &\cos \frac{\alpha}{2}|L\rangle+\sin \frac{\alpha}{2}|R\rangle\\
|-\rangle&=&\sin\frac{\alpha}{2}|L\rangle-\cos \frac{\alpha}{2}|R\rangle, \nonumber\\
\nonumber
\end{eqnarray}
with $\tan \alpha=\Delta_{0}/\Delta$. The last term in Eq.~(\ref{eq:smclassK}) denotes the field energy contained in the resonator
\begin{equation}
U=\frac{1}{2}\int d^{3}r\left[\epsilon F^{2}(t)+ \frac{B^{2}(t)}{\mu}\right],
\end{equation}
where $\epsilon$ and $\mu$ are the electric permittivity and magnetic 
susceptibility respectively and $B(t)$ is the magnetic field. The energy $U$ is usually not included in the classical model since the fields are not dynamical variables in that case; however it is relevant to the quantization of the problem below. 

\subsection{Quantum model}
To quantize the electromagnetic field in the Hamiltonian of Eq.~(\ref{eq:smclassK}) we use the prescriptions \cite{KnightBook}
\begin{equation}
\label{eq:pres}
U \rightarrow \hbar \omega a^{\dagger}a,\, F(t)\rightarrow F_{0}'\left[a(t)+a^{\dagger}(t)\right],
\end{equation}
where $\omega$ is the frequency of the resonator,
\begin{equation}
F_{0}'=\sqrt{\frac{\hbar \omega}{2\epsilon V}},
\end{equation}
is the `electric field per photon', $V$ is the effective mode volume of the resonator and $a(a^{\dagger})$ is an annihilation (creation)
operator obeying the bosonic commmutation rule $[a,a^{\dagger}]=1$.  
We thus obtain the quantum Hamiltonian,
\begin{eqnarray}
H_{Q}&=&\hbar \omega a^{\dagger}a+ \left[\frac{E}{2}
+p \cos \theta F_{0}'\frac{\Delta}{2E}(a+a^{\dagger})\right]\sigma_{z}\nonumber \\
&&+p \cos \theta F_{0}'\frac{\Delta_{0}}{2E}(a+a^{\dagger})(\sigma_{+}+\sigma_{-}),\\
\nonumber
\end{eqnarray}
where $\sigma_{\pm}$ are the TLS raising and lowering operators respectively.
Assuming the resonance condition $\hbar\omega=E,$ transforming to the 
interaction picture with respect to $H_{0}=\hbar \omega (a^{\dagger}a+ \frac{\sigma_{z}}{2})$ and making the rotating wave approximation, we arrive 
at
\begin{equation}
\label{eq:JCclass}
H_{Q}'=p \cos \theta F_{0}' \frac{\Delta_{0}}{2E}(a \sigma_{+}+a^{\dagger}\sigma_{-}),
\end{equation} 
which is of the form of the standard JCM, $H_{JCM}=\hbar g(a \sigma_{+}+a^{\dagger}\sigma_{-}),$ with a coupling constant given by
\begin{equation}
\label{eq:RabiK}
g=\frac{p\cos\theta F_{0}'\Delta_{0}}{2\hbar E}.
\end{equation}
In the remainder of the paper we will persist in using $g$ for the sake of avoiding lengthy expressions. Below we will consider the evolution of the TLS-resonator system when it is connected to reservoirs that cause relaxation and dephasing. 
\section{Equations of motion  : numerical treatment}
\label{sec:NUM}
In this section we will consider the full quantum mechanical treatment of the problem numerically via solution of the density matrix, and also present an approximate but simpler numerical approach using the equations for the expectation values of the relevant physical quantities.

\subsection{Master equation}
The presence of dissipation and dephasing in our problem can be accounted for by using the standard master equation approach for the JCM, which yields, in the Born-Markov approximation, an equation of motion for the density matrix $\rho$ of the TLS-resonator system \cite{Knight1991},
\begin{eqnarray}
\label{eq:MasterSHO2}
\dot{\rho}&=&-ig[a^{\dagger}\sigma_{-}+a\sigma_{+},\rho]
+\frac{1}{2T_{\phi}}(\sigma_{z}\rho\sigma_{z}-\rho)
\nonumber\\
&&+\frac{1}{2T_{1}}\left( 2\sigma_{-}\rho \sigma_{+}-\sigma_{+}\sigma_{-}\rho-\rho \sigma_{+}\sigma_{-}\right). \\
\nonumber
\end{eqnarray} 
The first term signifies unitary evolution, the second corresponds to
coupling to a reservoir that causes pure dephasing of the TLS at the rate $1/T_{\phi}$ and the last term denotes a coupling with a zero-temperature reservoir into which the TLS can relax at the rate $1/T_{1}$. 

Our approach will be to begin with the superconducting resonator in a coherent state $|\alpha \rangle$ (with average photon number $|\alpha|^{2}=\langle n(0)\rangle$) and the TLS in its ground state, and study the rate at which quanta are lost from the resonator. For simplicity, we have not included an external drive for the resonator. We have also assumed that the resonator suffers no intrinsic loss since we wish to study loss via the TLSs  \cite{Martinis2009,Osborn2010}. The typical frequencies of resonator operation correspond to energies much higher than available in a cryogenic environment $(\hbar \omega \gg k_{B}T)$, justifying our assumption of a zero-temperature reservoir above. Other regimes of the damped TLS-oscillator system have been addressed in, for example Ref.~\onlinecite{Milburn2008}.

The numerical solutions to Eq.~(\ref{eq:MasterSHO2}) will be discussed in the
context of the analytic results presented in Section ~\ref{sec:AN} below.

\subsection{Maxwell-Bloch equations}
While Eq.~(\ref{eq:MasterSHO2}) allows us to obtain a fully quantum mechanical
solution to the problem, it can involve the population and coherence dynamics
of a large number of states, especially for high initial photon numbers in the resonator. In this case it is useful to have a less intensive, and only slightly less rigorous, numerical approach to the problem. This route is provided by
the Maxwell-Bloch equations \cite{Kimble1998} which follow from Eq.~(\ref{eq:MasterSHO2}),
\begin{eqnarray}
\label{eq:MB1}
\langle \dot{a}\rangle&=&-ig\langle \sigma_{-}\rangle, \\
\label{eq:MB2}
\langle \dot{\sigma}_{-}\rangle&=&2ig \langle a \sigma_{++}\rangle-ig\langle a \rangle -\frac{1}{T_{2}}\langle \sigma_{-}\rangle,\\
\label{eq:MB3}
\langle \dot{\sigma}_{++}\rangle&=&-ig\left( \langle a \sigma_{+}\rangle-\langle a^{\dagger}\sigma_{-}\rangle\right)-\frac{1}{T_{1}} \langle \sigma_{++}\rangle, \\
\nonumber
\end{eqnarray}
where $\sigma_{++}=(1+\sigma_{z})/2$ is the population operator for the upper TLS level $|+\rangle$, and
\begin{equation}
\frac{1}{T_{2}}=\frac{1}{2T_{1}}+\frac{1}{T_{\phi}}.
\end{equation}
We can also find from Eq.~(\ref{eq:MasterSHO2}) the equation for the dynamics of the average photon number,
\begin{equation}
\label{eq:NUMMB}
\langle \dot{n}(t)\rangle=\frac{d}{dt}\langle a^{\dagger}a \rangle=ig\left( \langle a \sigma_{+}\rangle-\langle a^{\dagger}\sigma_{-}\rangle\right). 
\end{equation}
Combining Eqs.~(\ref{eq:MB3})  and (\ref{eq:NUMMB}) we find,
\begin{equation}
\label{eq:En}
\frac{d}{dt}\left[\langle n(t) \rangle + \langle \sigma_{++}\rangle \right]=-\frac{1}{T_{1}} \langle \sigma_{++}\rangle,
\end{equation}
which represents the conservation of energy in the system. The LHS of
Eq.~(\ref{eq:En}) denotes the rate of change of the the combined 
TLS-oscillator energy, while the RHS signifies the rate at which the TLS
releases quanta into the bath. We will use Eq.~(\ref{eq:En}) to generate 
simple analytical results below.

 It can be seen that Eqs.~(\ref{eq:MB1})-(\ref{eq:MB3}) do not constitute a closed set of equations. While the correlations in those equations are relevant to the full quantum solution of the problem, they turn out to play a negligible role in two cases : a) when the oscillator excitation is very low, since the TLS is then mostly in its ground state which then factors from the resonator state, and b) when the oscillator excitation is very high, since in this case the TLS is in a fully mixed state. We therefore assume decorrelations such as $\langle a \sigma_{+}\rangle \simeq \langle a \rangle \langle \sigma_{+}\rangle,$ etc.,
so that  Eqs.~(\ref{eq:MB1})-(\ref{eq:MB3}) now become
\begin{eqnarray}
\label{eq:MBD1}
\langle \dot{a}\rangle&=&-ig\langle \sigma_{-}\rangle, \\
\label{eq:MBD2}
\langle \dot{\sigma}_{-}\rangle&=&2ig \langle a \rangle \langle \sigma_{++}\rangle-ig\langle a \rangle -\frac{1}{T_{2}}\langle \sigma_{-}\rangle,\\
\label{eq:MBD3}
\langle \dot{\sigma}_{++}\rangle&=&-ig\left( \langle a \rangle \langle \sigma_{+}\rangle-\langle a^{\dagger}\rangle \langle \sigma_{-}\rangle\right)-\frac{1}{T_{1}} \langle \sigma_{++}\rangle, \\
\nonumber
\end{eqnarray}
which form a set of four closed equations if the conjugate of Eq.~(\ref{eq:MBD2}) is included. The numerical solutions to these equations are substantially quicker to obtain compared to the master equation and will be discussed below with reference to their approximate analytic solutions.

\section{Equations of motion : analytical treatment}
\label{sec:AN}
In order to  identify the various regimes of loss and to organize the numerical solutions to the master equation [Eq.~(\ref{eq:MasterSHO2})], we now discuss the problem in terms of approximate analytical solutions to the equations of motion. It is well known from previous studies that the JCM displays qualitatively different behavior in the regimes of weak $(g < 1/T_{1},1/T_{2})$ and strong $(g > 1/T_{1},1/T_{2})$ coupling \cite{Shore1993}. We now consider the two regimes separately.

\subsection{Weak coupling : $g < 1/T_{1},1/T_{2}$}
For weak coupling the JCM dynamics are dominated by relaxation and dephasing processes, allowing for little coherence between
the resonator and the TLS. Below we will see that in this case the TLS may be considered as being driven by the resonator at a time-dependent Rabi frequency 
\begin{equation}
\Omega_{q}=2g \langle n(t) \rangle^{1/2}.
\end{equation}
To find analytical solutions in this regime we consider the following argument.

If the Rabi drive due to the resonator could be held constant, we could obtain steady state solutions for the TLS variables. However the drive depends on the number of resonator quanta, and thus itself changes with time. Nonetheless, if the photon number changes slowly compared to the rate at which the TLS relaxes to equilibrium, we can assume the drive to be quasistatic, evaluate the TLS variables in the steady state, and then finally solve for the resonator dynamics. The above scheme amounts to a Born-Oppenheimer approximation where the TLS dynamics are considered as fast and the resonator dynamics as slow. 

Solving Eqs.~(\ref{eq:MBD1})-(\ref{eq:MBD3}) for the TLS variables in the quasistatic state we find the population in the upper TLS state $|+\rangle $ to be 
\begin{equation}
\label{eq:popTLS}
\langle \sigma_{++} \rangle= \frac{1}{2}\frac{R^{2}(t)}{1+R^{2}(t)},
\end{equation}
where 
\begin{equation}
\label{eq:RabiR}
R(t)=\Omega_{q} \sqrt{T_{1}T_{2}},
\end{equation}
is the Rabi frequency divided by the geometric mean of the longitudinal and transverse decay rates. Also 
\begin{equation}
|\langle \sigma_{+}\rangle|^{2}=\frac{T_{2}}{4T_{1}}\left[\frac{R(t)}{1+R^{2}(t)}\right]^{2},
\end{equation}
which shows that the coherence internal to the TLS is negligible at both small and large $R(t)$. The quantity $R(t)$, which is the ratio of the Rabi frequency to the geometric mean of the TLS decay rates, naturally delineates two qualitatively different types of behavior of the system, treated below.

\subsubsection{Unsaturated regime : $R(t) \ll 1$}
In this regime we find from the numerics that for initially small Rabi frequency, i.e. $R(0) \ll 1 $, the resonator photon number decreases exponentially; thus it is true that $R(t) \ll 1$
for all time $t$. From Eq.~(\ref{eq:popTLS}), we find
\begin{equation}
\label{eq:BOwet}
\langle \sigma_{++} \rangle \simeq \frac{R^{2}(t)}{2}.
\end{equation} 
Using this approximation we can solve Eq.~(\ref{eq:En}) to obtain
\begin{equation}
\label{eq:annum}
\langle n(t) \rangle =\langle n(0) \rangle e^{-\Gamma t},
\end{equation}
with the inverse of the effective decay rate given by
\begin{equation}
\label{eq:bigG}
\Gamma^{-1}=\frac{1}{2g^{2}T_{2}}+T_{1}.
\end{equation}
The first term on the RHS of  Eq.~(\ref{eq:bigG}) corresponds to the time required to transfer a single quantum from the resonator to the TLS. The 
second term corresponds to the time required for the TLS to release each quantum to the bath. The first time interval decreases with increasing coupling rate $g$ and the dephasing time $T_2$. However in the weak coupling regime this term is always larger than $T_{1}$.
 
\begin{figure}
\includegraphics[width=0.48 \textwidth]{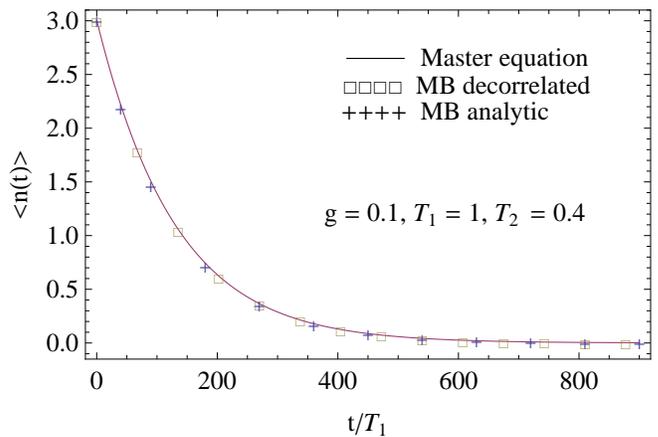}
\caption{This figure shows the time decay of the resonator photon number from the initial value $\langle n(0)\rangle =3$ in the weak coupling, TLS-unsaturated regime for $R(0) \simeq 0.2$. The three sets of points correspond to the numerical solution of the master equation, the numerical solution of the decorrelated Maxwell-Bloch equations and the analytical solution of Eq.~(\ref{eq:annum}), respectively, and are in quite good mutual agreement.}
\label{fig:WWN}
\end{figure}

As shown in Fig.~\ref{fig:WWN}
the analytical expression of Eq.~(\ref{eq:annum}) agrees well with the numerical solutions of both the master equation and the Maxwell-Bloch equations. We note that exponential decay of oscillator energy corresponds classically to the motion of 
a pendulum damped by a viscous fluid such as air, and is often referred to as
`wet' friction.

\begin{figure}  
\includegraphics[width=0.48 \textwidth]{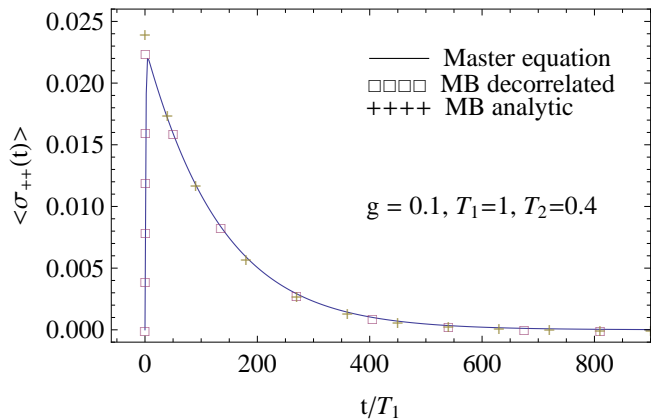}
\caption{This figure shows the dynamics of the population in the TLS level $|+\rangle$ starting from the initial value $\langle \sigma_{++}(0)\rangle =0$ in the weak coupling, TLS-unsaturated regime, for $R(0) \simeq 0.2$. The numerical solution to the master and decorrelated Maxwell-Bloch equations match quite well, whereas the analytical Born-Oppenheimer solution of Eq.~(\ref{eq:BOwet}) fails to capture the rapid initial excitation of the TLS.}
\label{fig:WWTLS}
\end{figure}

The analytical solution for the population $\langle \sigma_{++}(t)\rangle$ can be found self-consistently by using Eq.~(\ref{eq:annum}) in Eq.~(\ref{eq:BOwet}). The dynamics of the population are shown in Fig.~\ref{fig:WWTLS} and exhibit the validity of the analytical solution except for the non-adiabatic behavior at early times when the TLS undergoes rapid excitation in our numerical simulations.
It can be seen in this regime that the TLS is far from saturation and transports quanta efficiently into the reservoir. 

We note that our $R(t) \ll 1$ theory agrees with the result
for a qubit coupled to a TLS \cite{Makhlin2009}. The expression for the 
qubit population relaxation rate for weak coupling in Ref.~\onlinecite{Makhlin2009}, i.e. $v_{\perp}^{2}/\gamma_{1}^{f},$ agrees with the first and dominant term of Eq.~(\ref{eq:bigG}) with the appropriate substitutions $v_{\perp}=2g$ and $\gamma_{1}^{f}=1/T_{1}$ in the absence of pure dephasing $(T_{2}=2T_{1}).$

\subsubsection{Saturated regime : $R(t) \gg 1$}
\begin{figure}
\includegraphics[width=0.48 \textwidth]{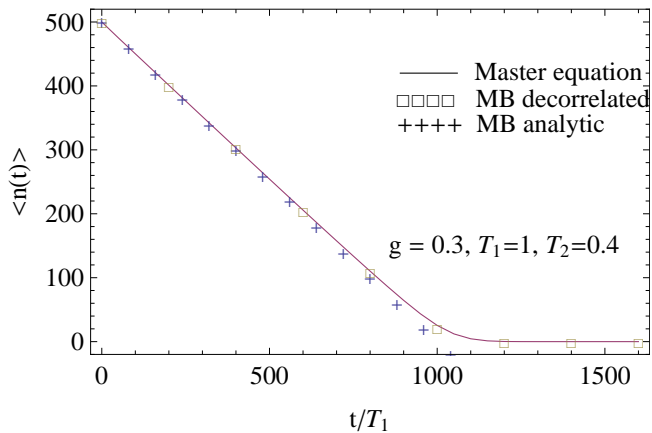}
\caption{This figure shows the time decay of photon number from the initial value $\langle n(0)\rangle =500$ in the weak coupling, TLS-saturated regime for $R(0)\simeq 4.2$. The three sets of points correspond to the numerical solution of the master equation, the numerical solution of the decorrelated Maxwell-Bloch equations and the analytical solution of
Eq.~(\ref{eq:annum2}), respectively. The first two agree well, while the analytical expression captures the behavior in the dominant linear regime.}
\label{fig:WDN}
\end{figure}

In this regime we find from the numerics that for initially large Rabi frequency, i.e. $R(0) \gg 1 $, the resonator photon number decreases essentially linearly with time. For this regime we find from Eq.~(\ref{eq:popTLS}), 
\begin{equation}
\label{eq:BOdry}
\langle \sigma_{++} \rangle \simeq 1/2.
\end{equation}
Using this approximation we can solve Eq.~(\ref{eq:En}) to obtain
\begin{equation}
\label{eq:annum2}
\langle n(t) \rangle =\langle n(0) \rangle -\frac{1}{2T_{1}}t.
\end{equation}
A plot of Eq.~(\ref{eq:annum2}) shown in Fig.~\ref{fig:WDN} 
matches well the numerical calculation in the regime of linear decay and gives correctly a slope of $-1/2T_{1}.$ The linearity of the decay is violated only at very long times when the photon number becomes very low and $R(t)$ is no longer large compared to unity.

We note that linear decay of oscillator energy corresponds classically to the motion of an oscillator impeded by a frictional force
always opposed to the oscillator's velocity but constant in magnitude, and is referred to as `dry' friction. An analytically solvable example is of a mass on a spring, vibrating horizontally on a surface and damped by sliding friction proportional to the normal force  \cite{Lapidus1970,Burko1999}. Clearly the frictional force `saturates' at a constant value in this instance. A comparison may be made between Fig.~\ref{fig:WDN} and that for the energy decay for a classical oscillator under dry friction (Fig.\ 5 in Ref.~\onlinecite{Burko1999}), including the nonlinear decay towards the end. 

\begin{figure}
\includegraphics[width=0.48 \textwidth]{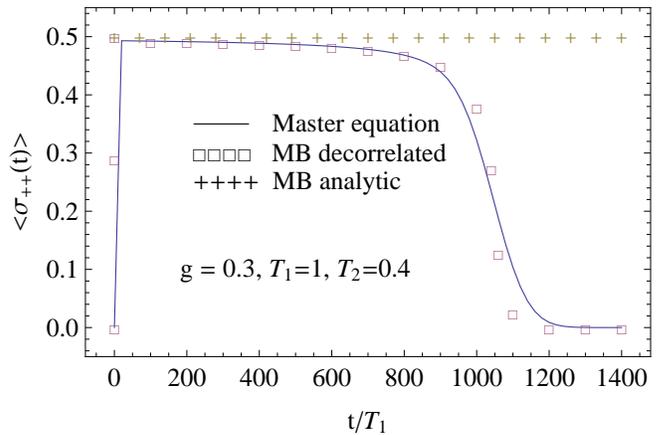}
\caption{This figure shows the dynamics of the population in the TLS  upper level  starting from the initial value $\langle \sigma_{++}(0)\rangle =0$ in the weak coupling, TLS-saturated regime, for $R(0)\simeq 4.2$. The numerical solution to the master and decorrelated Maxwell-Bloch equations match quite well. The analytical Born-Oppenheimer solution of Eq.~(\ref{eq:BOdry}) is constant and approximates the fact that the TLS is saturated most of the time.}
\label{fig:WDTLS}
\end{figure}

The population dynamics of the TLS are shown in Fig.~\ref{fig:WDTLS}.
 It can be seen that in this regime the TLS is almost always saturated and therefore can only accept quanta from the resonator limited by the rate at which it can release them into the bath. 

\subsubsection{Loss tangent}
We now consider the loss tangent, i.e. the behavior of the resonator loss
$1/Q$ where $Q$ is the quality factor, as a function
of $\langle n(0)\rangle,$ the initial average photon number. Typically modelling dielectric loss requires including the loss model for many TLSs, due to a distribution in $\Delta$ arising from the amorphous nature of the material.
This procedure is usually performed by averaging the solutions from the classical model and could be carried out numerically by extending the quantum model that we consider here.

To begin we recall the classical result, which follows from Eq.~(\ref{eq:smclassK})\cite{Phillips1987},
\begin{equation}
\label{eq:lossC}
\frac{1}{Q_{C}}=\frac{\hbar T_{2}}{2U}\left(\frac{\Omega^{2}}{1+\Omega^{2}T_{1}T_{2}}\right),
\end{equation}
with the classical Rabi frequency given by 
\begin{equation}
\label{eq:classRabi}
\Omega=\frac{p\cos\theta \Delta_{0} F_{0}}{2\hbar E},
\end{equation}
where $F_{0}(\neq F_{0}')$ is the amplitude of the classical electric field.
In this model saturation of the loss occurs at the critical Rabi frequency given by
$\Omega_{c}=(T_{1}T_{2})^{-1/2},$ which implies, via Eq.~(\ref{eq:classRabi}), a critical electric field 
\begin{equation}
F_{c}=\frac{2\hbar E}{p \cos \theta \Delta_{0}\sqrt{T_{1}T_{2}}}.
\end{equation}

For a proper comparison to the quantum results the classical field should be equated to the expectation value of
the quantum field operator [see Eq.~(\ref{eq:pres})] in the coherent state $|\alpha \rangle :$
\begin{equation}
F_{0} = \langle \alpha | F_{0}'(a+a^{\dagger})|\alpha \rangle =2\langle n(0)\rangle^{1/2}F_{0}'.
\end{equation} 
This yields, finally,
\begin{equation}
\label{eq:lossC2}
\frac{1}{Q_{C}}=\frac{2g^{2}T_{2}}{\omega [1+R^{2}(0)]}.
\end{equation}
The classical loss tangent is usually plotted as a function of the dimensionless electric field
\begin{equation}
\label{eq:dimlessF}
\frac{F_{0}}{F_{c}}=R(0).
\end{equation}
We note that the classical expression of Eq.~(\ref{eq:lossC2}) models a steady state measurement of the loss when the resonator is driven so as to ensure $\langle n(t)\rangle \equiv \langle n(0)\rangle$.

Now we turn to the results of the quantum theory, keeping in mind that there is no drive in our theory. When the TLS is unsaturated, the photon number decay is exponential, and the loss tangent 
may be defined as $\langle \dot{n} \rangle=-\omega \langle n \rangle/Q,$ or
\begin{equation}
Q \equiv -\omega\left.\frac{\langle n(t)\rangle}{\langle \dot{n}(t)\rangle}\right|_{t=0},
\end{equation}
where we have evaluated the quality factor at $\langle n(t=0)\rangle$ in order to make a valid comparison to the classical case.
This implies from Eq.~(\ref{eq:annum}) that
\begin{equation}
\label{eq:decayform}
\frac{1}{Q_{R \ll 1}}=\frac{\Gamma}{\omega}.
\end{equation}
For very weak coupling, $Q_{R \ll 1}^{-1}\simeq 2g^{2}T_{2}/\omega,$ which  reproduces the classical result of Eq.~(\ref{eq:lossC2}) with $R(0) \ll 1.$  

When the TLS is saturated the photon number decay is not exponential. In this case we find that the overall decay is typically dominated by the linear regime, especially for $\langle n(0)\rangle \gg 1$ (see  Fig.~\ref{fig:WDN}). Using Eq.~(\ref{eq:annum2}),
\begin{equation}
\label{eq:QQ}
\frac{1}{Q_{R \gg 1}}=\frac{1}{2 \langle n(0)\rangle T_{1}\omega},
\end{equation}
which agrees with the classical result of Eq.~(\ref{eq:lossC2}) for $R(0) \gg 1$, reproducing in particular the inverse scaling of loss with energy $[\propto \langle n(0) \rangle^{-1}]$. 

From the above analysis we may expect the onset of TLS saturation to occur at a critical number of quanta $n_{w}$ given by the intersection of the unsaturated [Eq.~(\ref{eq:decayform})] and saturated [Eq.~(\ref{eq:QQ})] loss asymptotes,
\begin{equation}
\label{eq:knee}
n_{w}=\frac{1}{2T_{1}\Gamma}\simeq \frac{1}{4 g^2 T_{1}T_{2}},
\end{equation}
where the second equality correponds to the very weak coupling regime and also to the condition $R=1$, in agreement with the classical theory. The quantity $n_{w}$ denotes the number of resonator photons required to saturate the TLS; in the weak coupling regime we expect $n_{w}>1.$

\begin{figure}[!] 
\includegraphics[width=0.5\textwidth]{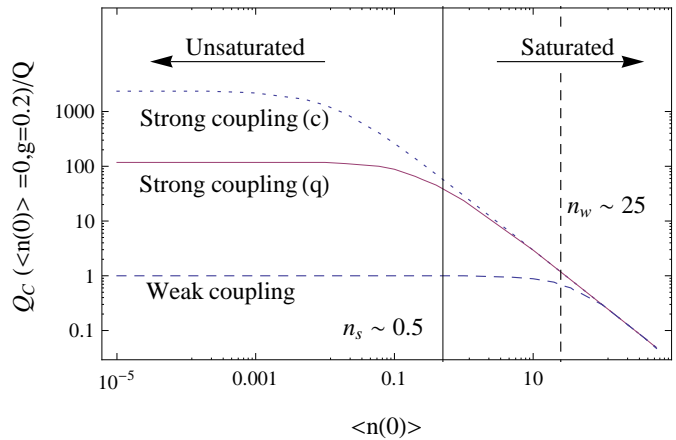}
\caption{This figure shows loss tangents of the resonator as a function of  initial photon number for various parameter regimes. The weak coupling quantum loss tangent (dashed curve) was generated using the values $g=0.2,T_{1}=1,T_{2}=0.2$. The strong coupling quantum and classical loss tangents (solid and dotted curves respectively) were both generated using the parameters $g=10,T_{1}=1,T_{2}=0.2.$ All curves are normalized to the weak coupling loss at low photon number given by Eq.~(\ref{eq:decayform}). As expected, quantum theory implies a higher loss for strong coupling of the resonator to the TLS than for the weak. However the loss according to quantum theory is much less than predicted by the classical theory using the strong coupling parameters. The photon number at which the loss saturates is $n_{w}(n_{s})$ for weak(strong) coupling and is shown as a dashed (solid) vertical line. }
\label{fig:QP}
\end{figure}

\begin{figure}[!] 
\includegraphics[width=0.5\textwidth]{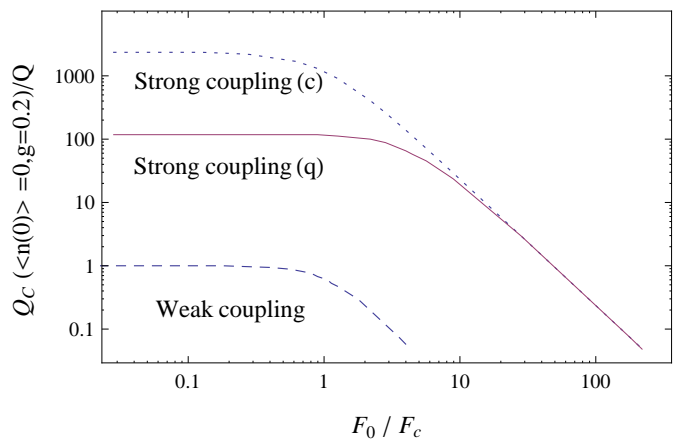}
\caption{This figure shows loss tangents of the resonator as a function of the 
dimensionless electric field of Eq.~(\ref{eq:dimlessF}) for various parameter regimes. The weak coupling quantum loss tangent (dashed curve) was generated using the values $g=0.2,T_{1}=1,T_{2}=0.2$. The strong coupling quantum and classical loss tangents (solid and dotted curves respectively) were generated using the parameters $g=10,T_{1}=1,T_{2}=0.2.$ All curves are normalized to the weak coupling loss at low photon number given by Eq.~(\ref{eq:decayform}). A comparison to Fig.~\ref{fig:QP} shows that while saturation occurs at a lower photon number for strong coupling compared to weak, it occurs at a larger electric field.}
\label{fig:QP2}
\end{figure}

Loss tangents were numerically calculated using Eq.~(\ref{eq:MasterSHO2}) for specific values of $g, T_{1}$ and $T_{2}$ and are shown in Fig.~\ref{fig:QP} (as a function of $\langle n(0)\rangle$) 
and  Fig.~\ref{fig:QP2} 
(as a function of $F_{0}/F_{c}$).

\subsection{Strong coupling : $g > 1/T_{1},1/T_{2}$}
In the strong coupling regime of the JCM, the coherent exchange of energy
between the resonator and the TLS plays an important role, although the 
overall dynamics is still governed by damping and dephasing.

\subsubsection{Unsaturated regime : $n \ll n_s$}
For small resonator photon numbers, we
notice from the numerics that the coherent energy transfer between the resonator and the TLS occurs at about the vacuum Rabi frequency $2g > (1/T_{1},1/T_{2})$. Thus the resonator photon number no longer changes slowly compared to the rate of TLS relaxation and the Born-Oppenheimer approximation employed above in the case of weak coupling ceases to be valid. 

However an analytic understanding may still be gained
from Eq.~(\ref{eq:MasterSHO2}) by observing from the numerics that for
low photon numbers the system dynamics is limited to a small effective Hilbert space. The space is defined by only three states : $|0,- \rangle, |1,-\rangle,$ and $|0,+\rangle,$ where the first entry in each ket denotes the resonator number state and the second the TLS state \cite{Kimble1998}. In this restricted manifold
noting that $\langle a^{\dagger}a \sigma_{z}\rangle \equiv -\langle a^{\dagger}a \rangle,$ we find from Eq.~(\ref{eq:MasterSHO2})
\begin{equation}
\label{eq:MasterAn}
\frac{d}{dt}\langle a^{\dagger}\sigma_{-}\rangle=ig (\langle\sigma_{++}\rangle-\langle n\rangle)-\frac{\langle a^{\dagger}\sigma_{-}\rangle}{T_{2}}.
\end{equation}
Together with its conjugate, Eq.~(\ref{eq:MasterAn}) forms a closed set of equations with Eqs.~(\ref{eq:MB3}) and (\ref{eq:NUMMB}). These equations can be solved subject to initial conditions set by the average photon number in the resonator $\langle n(0) \rangle$, the lack of TLS excitation $\langle \sigma_{++}(0)\rangle = 0,$ and the absence of correlations in the system $\langle a^{\dagger}\sigma_{-}(0)\rangle=0.$ 

As a simple example let us first completely neglect pure dephasing $(T_{2}=2T_{1}).$ We find, from Eqs.~(\ref{eq:MB3}) ,(\ref{eq:NUMMB}) and (\ref{eq:MasterAn}) the solution for the photon number,
\begin{equation}
\label{eq:VR1}
\langle n(t)\rangle=\langle n(0)\rangle e^{-t/2T_{1}}\cos^{2}gt,
\end{equation}
for the TLS population in the state $|+\rangle$,
\begin{equation}
\langle \sigma_{++}(t)\rangle=\langle n(0)\rangle e^{-t/2T_{1}}\sin^{2}gt,
\end{equation}
and for the correlation
\begin{equation}
|\langle a^{\dagger}\sigma_{-}(t)\rangle|^{2}=\langle n(0)\rangle^{2} e^{-t/2T_{1}}\sin^{2}2gt/4.
\end{equation}
We note that these results agree with those for strong coupling
between a qubit and a TLS \cite{Makhlin2009} as can be seen by using 
the substitutions $v_{\perp}=2g, \delta\omega=0$ and $\gamma_{1}^{f}=1/T_{1}$
in Eq. (4) of Reference \onlinecite{Makhlin2009}.

\begin{figure}
\includegraphics[width=0.48 \textwidth]{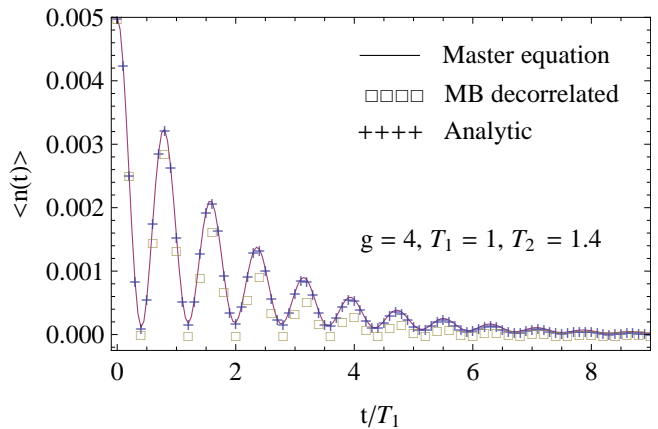}
\caption{This figure shows the photon number dynamics starting from the initial value $\langle n(0)\rangle =0.005$ in the strong coupling, TLS-unsaturated friction regime. The three sets of points correspond to the numerical solution of the master equation, the numerical solution of the decorrelated Maxwell-Bloch equations and the analytical solution to the system of equations given by Eqs.~(\ref{eq:MB3}), (\ref{eq:NUMMB}) and (\ref{eq:MasterAn}), respectively.}
\label{fig:SWN}
\end{figure}

\begin{figure}  
\includegraphics[width=0.48 \textwidth]{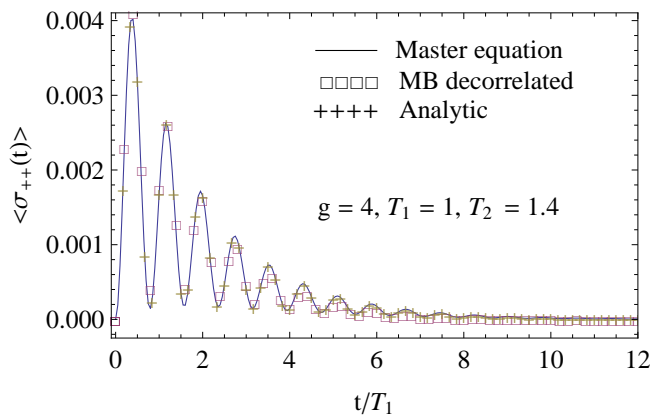}
\caption{This figure shows the dynamics of the population in the TLS level  starting from the initial value $\langle \sigma_{++}(0)\rangle =0$ in the strong coupling, TLS-unsaturated regime. The three sets of points correspond to the numerical solution of the master equation, the numerical solution of the decorrelated Maxwell-Bloch equations and the analytical solution to the system of equations given by Eqs.~(\ref{eq:MB3}), (\ref{eq:NUMMB}) and (\ref{eq:MasterAn}), respectively.}
\label{fig:SWTLS}
\end{figure}

In the presence of pure dephasing, Eqs.~(\ref{eq:MB3}), (\ref{eq:NUMMB}) and (\ref{eq:MasterAn}) can still be solved analytically but the results are quite cumbersome and we do not present the formulas here. They are instead plotted in Fig.~\ref{fig:SWN} 
and Fig.~\ref{fig:SWTLS} and compare well with the numerical solutions of the full master and Maxwell-Bloch equations. 
When pure dephasing is included, our results do not agree exactly
with the qubit-resonator predictions of reference \onlinecite{Makhlin2009}. 
The differences can be traced to our use of the rotating wave approximation (which discards terms from the equation of motion that oscillate at the
bare TLS frequency $\omega$ but retains terms that oscillate with
frequency 2g) and the use in Ref.~\onlinecite{Makhlin2009} of the secular approximation to the Bloch-Redfield equation (which also discards terms that
oscillate with frequency 2g).

It may be noted that the results of this section were obtained from the master
equation of Eq.~(\ref{eq:MasterSHO2}). That master equation was arrived at by
first considering a TLS and its dissipation. A lossless oscillator was then coupled to
the TLS. It has been pointed out that while this procedure yields a good approximation in the case of weak coupling, a different master equation ought to be used for the case of strong coupling, one derived by first diagonalizing the strongly coupled TLS and oscillator, and then adding dissipation to the whole   
system \cite{Walls1973,Murao1995,Murao1997,Zoubi2000}. We have verified that the resonator loss for strong coupling calculated from such a master equation agrees with that presented in this article for the TLS-unsaturated regime.

\subsubsection{Saturated regime : $n \gg n_s$}
We find from the numerics that for large photon numbers, the resonator
photon number decays linearly at a rate $1/2T_{1}$ and the TLS is saturated.
The resonator and TLS dynamics are similar to the case of weak coupling as shown in Figs.~\ref{fig:WDN} and \ref{fig:WDTLS}  respectively, and are well described by Eq.~(\ref{eq:annum2}) and Eq.~(\ref{eq:BOdry}) respectively. 

However we note two aspects with respect to which the strong coupling dynamics differs from that in the weak coupling case. First, superposed on the overall decay of photon number and saturation of the TLS are oscillations of small amplitude which occur at early times as coherent transients, and later as revivals familiar to the JCM \cite{Shore1993}. Second, saturation of the TLS does not occur at $R(t) \simeq 1$, in contrast to the case of weak coupling as discussed below.
 
\subsubsection{Loss tangent}
Although the qualitative shape of the loss tangent is similar to that of weak coupling, as can be seen from Fig.~\ref{fig:QP} and Fig.~\ref{fig:QP2} there are crucial physical differences between the two. First, the asymptotic value for strong coupling when the TLS is unsaturated is found from the analytic solution of Eqs.~(\ref{eq:MB3}), (\ref{eq:NUMMB}) and (\ref{eq:MasterAn}) to be 
\begin{equation}
\label{eq:losststrongwet}
\frac{1}{Q_{n\ll n_{s}}} = \frac{1}{3\omega}
\left(\frac{1}{T_{1}}+\frac{1}{T_{2}}\right)=\frac{1}{\omega}
\left(\frac{1}{2T_{1}}+\frac{1}{3T_{\phi}}\right).
\end{equation}
Our prediction, Eq.~(\ref{eq:losststrongwet}), indicates that for large $g$, the 
classical theory overestimates the loss in the unsaturated regime. Furthermore the result of Eq.~(\ref{eq:losststrongwet}) can be used to show straightforwardly that the quantum theory predicts a much greater loss in the unsaturated regime for large $g$ (strong coupling) than for small $g$ (weak coupling). Physically, this is reasonable -- loss is higher for the resonator when it is strongly coupled to the TLS. Comparing Eq.~(\ref{eq:losststrongwet}) to Eq.~(\ref{eq:bigG}) we see that the pure dephasing plays opposite roles in the two cases : loss in the weak coupling case decreases with pure dephasing while it increases for strong coupling.  In the saturated-regime all theories coincide, and follow Eq.~(\ref{eq:QQ}), which holds for strong as well as weak coupling. The loss curves are shown as a function of the dimensionless electric field of Eq.~(\ref{eq:dimlessF}) in Fig.~\ref{fig:QP2},
from which it can be seen that although the saturation photon number 
for strong coupling is lower than for weak coupling (Fig.~\ref{fig:QP}), the 
critical electric field at which saturation occurs is higher (Fig.~\ref{fig:QP2}).

Lastly, from the above considerations the crossover between the two regimes can be estimated to occur at a critical photon number $n_{s}$ given by the intersection of the unsaturated 
[Eq.~(\ref{eq:losststrongwet})] and saturated [Eq.~(\ref{eq:QQ})] loss asymptotes,
\begin{equation}
\label{eq:strongk}
n_{s}= \frac{3}{2}\left(1+\frac{T_{1}}{T_{2}}\right)^{-1}.
\end{equation}
We note that Eq.~(\ref{eq:strongk}) is valid for very strong coupling,
$g \gg 1/T_{1},1/T_{2}.$ This explains the absence of $g$ from Eq.~(\ref{eq:strongk}) : deep in the strong coupling regime $g$ is responsible for the reversible dynamics, while loss arises from irreversible processes. This argument also implies that the dimensionless quantity $n_s$ can then only depend on the ratio $T_{1}/T_{2}$ as verified by the final formula; for a large $T_{1}$ the resonator easily saturates the TLS and a small $T_{2}$ implies that the coherent oscillations via which the TLS returns quanta to the resonator are quickly dephased.  In the absence of pure dephasing $(T_{2}=2T_{1})$ saturation begins to turn on at $n_s =1$; for $T_{2} < 2T_{1}, n_{s} < 1$. However, since 
it is difficult to transiently saturate the TLS with $\langle n(0)\rangle \ll 1,$
$n_s$ does not take arbitrarily low values. A numerical search yields the bound
\begin{equation}
n_{s,\mathrm{min}} \geq 1/2.
\end{equation}

\section{Conclusion}
\label{sec:CONC}
We have quantized the classical model for dielectric loss in superconducting 
resonators due to two level charge fluctuators, arriving thus at the Jaynes-Cummings model of quantum optics. For both the strong and weak coupling scenarios of this model we have identified regimes corresponding to nonsaturation and saturation respectively of the TLS. We have found that the quantum theory agrees with the classical in the regime of weak coupling. However for strong coupling, we find the quantum theory prescribes a substantially greater loss than the classical. Moreover, the photon number at which TLS saturation occurs is independent of the resonator-TLS coupling in the strong coupling quantum theory, in complete contrast to the classical theory.

The numerical and analytical results presented in this paper can be used as a starting point for including further parameters in the loss model in order to make it more realistic, such as an external drive for the resonator, intrinsic resonator loss, and reservoirs at non-zero temperature. In addition, we have in this work only considered the situation where the resonator is resonant with the TLS.
The case of non-zero detuning is of interest since experiments typically include
a distribution of many TLSs with varying detunings. Usually distributions in the variables of orientation $(\theta)$, energy $(\Delta)$ and tunneling barriers $(\Delta_{0})$ also need to be considered in order to reach agreement with experiment \cite{Osborn2010}. Clearly there can be several additional regimes of behavior of the quantum loss model, depending on the relative values of the parameters introduced. While working in the time domain serves to illustrate
the dynamics clearly, working in the frequency domain can yield additional insights, and is relevant to some experiments; such a project is envisioned for the future. We would like to thank F. Wellstood, M. Stoutimore, and M. Khalil for discussions.

\end{document}